\documentclass[12pt]{article}
\usepackage{amsmath}
\usepackage{amssymb}
\usepackage{amsthm}
\usepackage{psfrag}
\usepackage{graphicx}
\usepackage{hyperref}

\newcommand{\be}{\begin{equation}}
\newcommand{\ee}{\end{equation}}
\newcommand{\beqa}{\begin{eqnarray}}
\newcommand{\eeqa}{\end{eqnarray}}

\newcommand\m{\mu}
\newcommand\D{\Delta}
\newcommand\n{\nu}
\renewcommand\r{\rho}
\newcommand\s{\sigma}
\renewcommand\a{\alpha}
\renewcommand\b{\beta}
\renewcommand\l{\lambda}

\def\e{{\rm e}}
\def\d{\partial}
\newcommand{\bseq}{\begin{subequations}}
\newcommand{\eseq}{\end{subequations}}

\renewcommand{\ln}{\mathop{\rm ln}\nolimits}

\renewcommand{\Im}{\mathop{\rm Im}\nolimits}

\newcommand{\di}{\mathrm d}

\title{
\sc{\huge
\vspace{-.35cm} Ho\v rava gravity vs. thermodynamics: the black hole case}
\date{}}
\author{D. Blas\,$^{a,b},$
  S. Sibiryakov\,$^{c},$\vspace{.1cm}\\
\normalsize\llap{$^a$}
 \it FSB/ITP/LPPC,
 \'Ecole Polytechnique F\'ed\'erale de Lausanne,\\
 \normalsize\it CH-1015, Lausanne, Switzerland\\
\normalsize\llap{$^b$} \it Center for Cosmology and Particle Physics,\\
\normalsize\it Department of Physics, New York University, New York, NY, 10003, USA\\
\normalsize\llap{$^c$} 
\it Institute for Nuclear Research of the
Russian Academy of Sciences, \\ 
      \normalsize \it  60th October Anniversary Prospect, 7a, 117312
      Moscow, Russia
\vspace{-.25cm}} 
\begin{document}

\maketitle

\begin{abstract}
Under broad assumptions breaking of Lorentz invariance in
gravitational theories leads to tension with unitarity because it allows for
processes that apparently violate the second law of
thermodynamics. The crucial ingredient of this argument is the existence of
black hole solutions with the interior shielded from infinity by a
causal horizon. We study  how the  paradox can be resolved in the
healthy extension of Ho\v rava gravity. To this aim  
we analyze classical solutions describing large black holes in this theory
with the emphasis on their
causal structure. The notion of
causality is subtle in this theory  due to the presence of
instantaneous interactions. Despite this fact, we find that
within exact spherical symmetry a black hole solution contains
a space-time region causally disconnected from infinity by a surface
of finite area -- the
`universal horizon'. We then 
consider small perturbations of arbitrary angular dependence
in the black hole background. We argue
that aspherical perturbations destabilize the universal horizon and,
at non-linear level,
turn it into a finite-area singularity. 
The causal structure of the region outside
the singularity is trivial. If the 
higher-derivative terms in the equations of motion 
smear the singularity 
while preserving the trivial causal structure of the
solutions, the thermodynamics paradox would be obviated. 
As a byproduct of our 
analysis we prove that the black holes do not have any non-standard
long-range hair. We also  comment on the 
relation with Einstein-aether theory, where the solutions with 
universal horizon appear to be stable.
\end{abstract} 

\newpage

\section{Introduction and summary}

There are several reasons for the current interest in gravitational models with broken Lorentz invariance (LI).
 First, these theories may be relevant in the quest for consistent alternatives to  general relativity (GR) that modify the laws of gravity at
large distances. (This search is in turn prompted by the
attempts to resolve the problems of dark matter and dark
energy raised by cosmological data.) Second, the construction of
these models 
provides a necessary framework for
testing the nature of LI. These motivations are behind the Einstein-aether model 
\cite{Jacobson:2000xp,Jacobson:2008aj}, where Lorentz breaking is
realized by a unit time-like vector field, and the ghost condensation
model 
\cite{ArkaniHamed:2003uy}, where LI is broken by a scalar field with
time-dependent vacuum expectation value (VEV). 
More
complicated patterns of Lorentz breaking are realized in
models of massive gravity \cite{Dubovsky:2004sg}, see
\cite{Rubakov:2008nh} for a review.\footnote{An important
  issue in theories with broken LI is to explain why Lorentz violation
does not propagate into the Standard Model sector of particle
physics where LI is tested with extreme accuracy. It has been
proposed  that LI may be protected by supersymmetry \cite{GrootNibbelink:2004za}. In
\cite{Pujolas:2011sk} this protection mechanism has been realized in
the context of 
the supersymmetric extension of the Einstein-aether model.}

More recently, this interest has been spurred by Ho\v rava's proposal \cite{Horava:2009uw} 
to construct a renormalizable model of quantum gravity
by giving up LI. The main idea is that in the absence
of LI the ultraviolet (UV) behavior of  gravitational amplitudes
can be substantially improved by the addition of terms with higher
spatial derivatives to the action. This can be done while keeping the
Lagrangian second order in time derivatives, thus avoiding problems
with unstable degrees of freedom appearing in LI
higher-derivative gravity \cite{Stelle:1977ry}.  The consistent implementation of this idea
involves the notion of anisotropic (Lifshitz) scaling of the theory in
the UV and indeed leads to a theory which is renormalizable by
power-counting \cite{Horava:2009uw}. The
renormalizability of the theory in the rigorous sense remains an open problem. 

Lorentz violating gravitational models 
generically contain new degrees of freedom besides the two
polarizations of the graviton. These degrees of freedom survive down to 
the infrared where they can be conveniently described as a new Lorentz
violating sector interacting with  
 Einstein's
general relativity. In this paper we will mainly focus on  
 Ho\v rava gravity. This model includes 
one new degree of freedom described by a scalar field $\varphi(x)$ with non-zero
time-like gradient 
\cite{Blas:2009yd,Blas:2010hb},\footnote{We use the metric signature $(+,-,-,-)$.}
\be
\label{grad}
\d_\mu\varphi\neq 0~,~~~~
(\d_\mu\varphi)^2>0\;.
\ee    
Because of the latter property $\varphi$ can be chosen as a time
coordinate and thus acquires the physical meaning of 
universal time; we will refer to it as `khronon'. The theory contains a mass scale
$M_*$ assumed to be just a few orders
of magnitude below the Planck mass and which suppresses the operators with higher derivatives.
 These operators are not important for the description of the khronon
 dynamics 
and its interaction with gravity at low energy. 
The low-energy properties are captured by an effective 'khronometric'
theory -- a scalar-tensor theory satisfying certain symmetries 
\cite{Blas:2010hb}. 
It is worth mentioning that, as it is always the case for effective
theories, the khronometric model is in a sense more general than 
Ho\v rava gravity: given the symmetries and field content, it can be
derived independently as the theory with the smallest number of
derivatives in the action.
In this paper we concentrate on the
`healthy' model proposed in 
\cite{Blas:2009qj,Blas:2009ck} which corresponds to the so called
`non-projectable' version of Ho\v rava gravity. In this case the
khronometric model possesses a symmetry under 
reparameterization of the khronon
field,
\be
\label{reparam}
\varphi\mapsto f(\varphi)\;,
\ee
where $f$ is an arbitrary monotonic function. Due to this symmetry the
model differs essentially from ghost condensation, despite the
fact that both are scalar-tensor theories with Lorentz 
violation\footnote{A model similar to the ghost condensation arises in
  the `projectable'' version of Ho\v rava gravity
  \cite{Blas:2009yd,Blas:2010hb}. However, in this case  the khronon
  develops instability or strong coupling in the Minkowski
  background.}.  
Instead, the model turns out to have a lot of
similarities with the Einstein-aether theory
\cite{Blas:2010hb,Jacobson:2010mx}, without, however, being completely
equivalent to it. Indeed, in addition to the scalar mode the aether 
contains two more propagating degrees of freedom corresponding to the
transverse polarization of a vector field. 
On the other hand, the khronometric model includes instantaneous
interactions \cite{Blas:2010hb} that are
absent in the case of Einstein-aether\footnote{This issue will be
  discussed in detail in Sec.~\ref{sec:2} and Appendix~\ref{App:B}.}.
The phenomenological consequences of the khronometric model have
been analyzed in \cite{Blas:2010hb,Blas:2011zd}, where it was shown 
that for appropriate choice of parameters 
it satisfies the existing experimental constraints\footnote{The emission 
of gravitational waves by binary systems containing sources with large self-energies has not yet been computed. However, 
for the systems where the radiation damping has been observed
 no big differences  with respect to the computations of the weak fields regime are
expected \cite{Blas:2011zd}.}. In \cite{Blas:2011en}
it was shown that extending the model by an additional scalar field
with exact shift symmetry allows to naturally account for the cosmological
dark energy.

In this paper we study black hole (BH) solutions of the khronometric
theory. These provide large distance (with respect to the scale $M_*^{-1}$) solutions of the
healthy Ho\v rava gravity, and are expected to be produced by
gravitational collapse. One expects BHs in Ho\v rava gravity to 
differ from those of GR  in several
aspects. 
In GR a BH
is characterized by the existence of a horizon for light rays, 
which prevents the
propagation of signals from its interior to the exterior. 
In this way, the horizon shields the central
singularity\footnote{It should be stressed that by 
  singularity we always understand singularity from the point of view of the
low-energy theory. In the full theory of quantum gravity the
singularity is expected to be resolved by the effects that become
important at high curvature (such as
stringy effects in string theory or higher-derivative terms in Ho\v
rava gravity).} of the BH from the exterior. However, in the
presence of Lorentz violation the theory may contain excitations whose
propagation velocity exceeds that of light and which thus can escape
from inside  GR horizons
\cite{Babichev:2006vx}. Moreover, in Ho\v rava gravity the dispersion
relations of the propagating degrees of freedom in the locally flat
coordinate system have the form \cite{Horava:2009uw}
\be
E^2=c^2_n p^2+\frac{a_n p^4}{M_*^2}+\frac{b_n p^6}{M_*^4}\;,
\ee
where $E$ and $p$ are the energy of the particle and its spatial
momentum, and $c_n$, $a_n$, $b_n$ are coefficients of order one
depending on the particle species $n$. Stability at high momenta
requires the coefficient $b_n$ to be positive.
This implies that both phase
and group velocities of particles indefinitely grow with energy
and one might think that these modes can come from the immediate
vicinity of the central singularity. Finally, as we discuss in
Sec.~\ref{sec:2}, even within the low-energy description in terms of
the khronometric model the theory contains a certain type of
instantaneous interactions, which again might probe the BH interior
down to the center. This suggests that Ho\v rava gravity does not
allow for BHs in the strict sense, characterized by the existence of
regions causally disconnected from the asymptotic infinity. The
purpose of this article is to clarify if this expectation is true.

In our study we will be guided by the puzzles of BH thermodynamics
arising in theories with Lorentz violation.
As pointed out in \cite{Dubovsky:2006vk,Eling:2007qd}
for the examples of
ghost condensation and Einstein-aether theories, 
one can construct gedanken experiments 
involving BHs that
violate the second law of thermodynamics (see, however,
\cite{Mukohyama:2009um} for a different point of view). 
Specifically, it is possible
to set up processes that would decrease the entropy in the region
outside the BH without any apparent change of the state of the BH itself.
The second law of thermodynamics is intimately related to the
unitarity of the underlying microscopic theory, see
e.g. \cite{weinberg}. Thus its violation would constitute a serious
problem, especially for a theory that, like Ho\v rava gravity, aims
at providing the microscopic description of quantum gravity. One can
consider the following scenarios to recover the second law of thermodynamics:
\begin{itemize}
\item[(i)]
The missing entropy is accumulated somewhere inside the BH. 
The first guess for the precise location of the entropy storage
region is close to the central singularity. In fact, this singularity
is expected to be 
smeared off in the full theory at distances set by the microscopic
scale and the entropy, in principle, can be stored in some high
frequency modes localized in this smeared region.
For this mechanism to work, an outside observer 
must be able to probe this region in order to make sure that the total
entropy of the system does not decrease. This option seems plausible,
given that Ho\v rava gravity contains 
arbitrarily fast modes including an instantaneous mode that persists 
at low energies. However, one
must check that these modes can indeed escape form the center of the
BH. In other words, one has to work out the causal structure of
 BHs in this theory. 

\item[(ii)]
Another possible scenario to restore the thermodynamics  assumes that
BHs are not
uniquely characterized by their mass, but instead can come out in many
different configurations \cite{Dubovsky:2007zi}. 
 In other words, it  supposes that BHs have a large
number of static
long-range hair. This hair would grow during the processes suggested in
\cite{Dubovsky:2006vk,Eling:2007qd} and after
measuring them an outer observer could
decode the entropy fallen down into the BH. Notice that the difference
with respect 
to the previous point is that the hair has a tail that can be measured
outside the horizon.  
It has been demonstrated for several Lorentz violating theories including
the Einstein-aether \cite{Eling:2006ec}, ghost condensate
\cite{Mukohyama:2005rw} and massive gravity \cite{Dubovsky:2007zi}
that spherically symmetric solutions with given mass are unique. 
As we
are going to see, this is also the case for the khronometric
model.
This implies that the hair necessarily must be non-spherical. 
\end{itemize}

One of the purposes of this paper is to clarify which of these scenarios, if any,
is realized in the healthy Ho\v rava
gravity. For this purpose,  we first find spherical BH solutions in the
khronometric model and analyze their causal structure. To simplify the
 analysis we will neglect the back-reaction of the khronon
field on the metric. This approximation, valid when the
dimensionless parameters of the khronon action are small,  reduces
the problem of finding the BH solution of the khronometric
model to that of embedding the khronon field into a given
metric background\footnote{A similar approach was used in 
\cite{Dubovsky:2007zi} for the analysis of BHs in massive
gravity.}. We 
show in passing that these solutions also describe spherical 
BHs in the Einstein-aether
theory.  
Next we
consider perturbations on top of these solutions. We demonstrate that no static
long-range hair exists thus
rejecting option (ii). On the other hand, option (i) is likely
to work, though in a quite non-trivial manner. Within spherical
symmetry we find that, despite the presence of arbitrarily fast,
viz. instantaneous, interactions, the center of the BH is
shielded by a causal horizon. This `universal horizon' lies inside
the Schwarzschild radius and its size linearly depends on the BH 
mass. All fields of the spherically symmetric solution are
regular (analytic) at this horizon. However, in the spectrum of
{\em non-spherical time-dependent} perturbations one finds certain
modes with non-analytic structure at the universal horizon. These
modes precisely correspond to the instantaneous interactions of the
khronometric theory. Though at the linear order the universal
horizon turns out to be stable under perturbations, we argue that the
above non-analyticities will destabilize it at non-linear level,
turning it into a finite-area singularity. One hopes that in the full
Ho\v rava gravity this singularity is resolved into a high-curvature
region of finite width accessible to the instantaneous and fast
high-energy modes. In this way thermodynamics can be saved.

It is worth stressing that the presence of instantaneous interaction is
crucial for the type of instability discussed in this
paper. Consequently, the universal horizon is expected to be stable in
the Einstein-aether theory where all modes propagate with finite
velocities. At present we are unable to suggest any resolution of the paradox with
BH thermodynamics in this theory. 

Despite the vast literature on BHs in Ho\v rava gravity, 
there has been only a few works dealing with the healthy extension in which we
are interested. A class of  spherically symmetric solutions of the healthy
Ho\v rava gravity was found in \cite{Kiritsis:2009vz}. Those solutions
differ, however, from the black holes considered in this paper. 
Ref.~\cite{Ted} has obtained spherically symmetric BHs in the
Einstein-aether and khronometric theories
by numerically solving the
coupled system of equations for the metric and aether (khronon)
field\footnote{For previous studies of BHs in the
  Einstein-aether model see \cite{Eling:2006ec,Garfinkle:2007bk}.}. 
Our approach is different in that we simplify the setup  to use analytic
techniques whenever possible. This allows us
to go beyond the spherical symmetry and study aspherical perturbations
around BHs. 
 Where our results overlap
with those of \cite{Ted}, they agree.

The paper is organized as follows. In Sec.~\ref{sec:2} we introduce
the khronometric model and demonstrate that it contains instantaneous
interactions. 
In Sec.~\ref{sec:3} we find spherical BHs and show that those are also
solutions for Einstein-aether theory. 
We also analyze their causal structure in this section. We turn to the
analysis of the khronon 
perturbations about the BH in Sec.~\ref{sec:hair}. 
Section \ref{sec:discussion} is devoted to 
discussion. In Appendix~\ref{App:B} we show that the
instantaneous interaction is absent in the Einstein-aether model.  
Appendix \ref{App:A} contains certain analytic results
about spherical BHs. 

\section{The khronometric model and  instantaneous modes}
\label{sec:2}

The khronometric action corresponding to the low-energy limit of Ho\v
rava gravity has the form 
\cite{Blas:2010hb},
\be
\label{Act}
S=-\frac{M^2}{2}\int \di^4x \sqrt{-g}\bigg[R+\a (u^\m\nabla_\m u_\n)^2
+\b \nabla_\m u^\n\nabla_\n u^\m + \l (\nabla_\m u^\m)^2\bigg]\;.
\ee
Here $R$ is the Ricci scalar and the unit vector $u_\m$ is expressed
in terms of the khronon field $\varphi$ as,
\be
\label{uphi}
u_\mu\equiv\frac{\d_\m\varphi}{\sqrt{\d_\n\varphi\nabla^\n\varphi}}\;.
\ee
In Eq.~(\ref{Act}) $M$ is a mass parameter related to the Planck mass  and $\a,\b,\l$ are dimensionless
constants\footnote{The parameter $\lambda$ in (\ref{Act}) corresponds
to $\lambda'$ in the notations of \cite{Blas:2010hb}.}. 
The previous action is the most
general expression invariant under the
transformations (\ref{reparam}) and containing only two derivatives of
$u_\m$. Note that it formally coincides with the action of
the Einstein-aether theory\footnote{To be precise, the action of the
  Einstein-aether model contains an additional term proportional to
  $\nabla_\m u_\n\nabla^\m u^\n$. In the case of
  hypersurface-orthogonal aether (\ref{uphi}) this term is equal to a
  combination of 
 the terms already present in the action (\ref{Act}).}  
\cite{Jacobson:2000xp,Jacobson:2008aj}. The difference, however, is
that in the case of Einstein-aether the unit vector $u_\m$ is treated
as a fully dynamical variable, subject only 
to the constraint of been a unit time-like vector, 
and thus contains additional transverse
degrees of freedom compared to the khronometric model where it is written in terms
of the scalar field  $\varphi$.   
This difference disappears for spherically symmetric solutions, since
the aether vector 
is always hypersurface-orthogonal in that case \cite{Eling:2006ec}
and thus can be expressed in the form (\ref{uphi}). 

We assume that the parameters $\a,\b,\l$ are
small,
\be
\label{small}
\a,~\b,~\l\ll 1.
\ee
On one hand, this choice is motivated by the phenomenological
bounds on the model \cite{Blas:2010hb,Blas:2011zd}. At the same time
it drastically simplifies the rest of the analysis allowing to 
neglect the
back-reaction of the khronon field on the metric. In fact, it is clear
from (\ref{Act}) that the energy-momentum tensor of the khronon field is
proportional to the parameters $\a,\b,\l$ . Thus its contribution into
the Einstein equations is negligible when these parameters satisfy
(\ref{small}). 
This implies that the problem of finding a solution of the theory
 is reduced to solving the khronon equation of motion in an
external (background) metric.

A crucial property of the theory at hand is the presence of an instantaneous interaction
\cite{Blas:2010hb}. 
To understand its origin one can consider the
propagation of small khronon perturbations in Minkowski space-time. 
It is straightforward to see that the Ansatz $\varphi=t$ for the khronon
background satisfies the equations of motion in this space-time. For the
perturbations, we define a field $\chi$ on top of this background,
\be
\label{khpert}
\varphi=t+\chi\;,
\ee
for which one obtains the following quadratic action:
\be
\label{chiAct}
S^{(2)}_\varphi=\frac{M^2}{2}\int \di^4 x~\Big[\a (\d_i\dot\chi)^2
-(\b+\l)(\Delta\chi)^2\Big]\;.
\ee
Note that this action is fourth order in derivatives. However, the additional
derivatives are purely space-like and the action describes a single
wave mode with  velocity
\be
\label{soundc}
c_\chi=\sqrt{\frac{\b+\l}{\a}}\;.
\ee
As shown in \cite{Blas:2010hb}, this remains true also for curved
backgrounds: the equation of motion for khronon perturbations always
remain second order in time derivatives if the time coordinate is
chosen to coincide with the background khronon field. 

Still the presence of extra spatial derivatives implies
instantaneous propagation of signals. To see this, let us
 couple the khronon field to a source. The simplest
source term that preserves the symmetries of the khronon field has 
the form
\be
\label{Ssource}
S_{source}=\int \di^4 x\sqrt{-g}\, S^\m u_\m \;,
\ee
where $S^\m$ is the source vector. To linear order around the flat background this reads
\be
\label{Ssourcel}
S_{source}^{(1)}=\int \di^4x\;S^i\d_i\chi\;,
\ee
which leads to the khronon exchange amplitude
\be
\label{ampl}
{\cal A}=M^{-2}S^i(x)G_{ij}(x-y)S^j(y)\;,
\ee
where 
\be
\label{Green1}
G_{ij}(x)=\int\frac{\di^4p}{(2\pi)^4}
\frac{{\bf p}_i{\bf p}_j~\e^{-ipx}}{{\bf p}^2
\big(\a p_0^2-(\b+\l){\bf p^2}\big)}\;,
\ee
where $\theta$ is the step function and $\delta$ the Dirac
delta-function. 
A straightforward
calculation yields the retarded propagator:
\be
\label{Green}
G_{ij}^{ret}(x)=-\frac{\theta(t)\delta(|{\bf x}|-c_\chi t)}{4\pi\sqrt{\a(\b+\l)}}\,
\frac{{\bf x}_i{\bf x}_j}{|{\bf x}|^3}\,
+\frac{\theta(t)\theta(|{\bf x}|-c_\chi t)}{4\pi\a}\,
\left(\frac{3{\bf x}_i{\bf x}_j-\delta_{ij}|{\bf x}|^2}{|{\bf x}|^5}\right)t\;.
\ee
Clearly, due to the second
term the retarded Green's function does not vanish outside the khronon
`sound cone'\footnote{Curiously, the Green's function does vanish
  {\it inside} the `sound cone', as in the case of the ordinary  massless field
in four dimensions.} $c_\chi t=|{\bf x}|$. Moreover, the field extends
to arbitrary
spatial distance from the source immediately after the source is
switched on, meaning that the signal propagates instantaneously. 
Note, however, that the instantaneous part of the signal builds up
gradually starting from zero at $t=0$ and the maximum amplitude that is
reached by the time $t=|{\bf x}|/c_\chi$ decreases with the distance
as $1/|{\bf x}|^2$. This is similar to the situation in Lorentz
violating massive electrodynamics \cite{Gab,Dvali:2005nt} and massive
gravity \cite{Dubovsky:2007zi,Bebronne:2008tr}.
It is worth stressing that due to the existence of a preferred frame
the instantaneous propagation of 
signals does not lead to any inconsistencies.  

Note that the instantaneous piece in (\ref{Green}) is
traceless and transverse,
\[
G_{ii}^{ret,\,inst}(x)=\d_i G_{ij}^{ret,\,inst}(x)=0\;.
\]
The latter property implies that the instantaneous contribution will
vanish if the source $S^\m(x)$ is a gradient of a localized scalar
configuration. We also point out that, despite the similarity between
the khronometric and Einstein-aether theories, the latter does not
contain any instantaneous interactions. As shown in the
Appendix~\ref{App:B}, the instantaneous piece in (\ref{Green}) is
canceled by the contributions of the transverse modes.

\section{Spherical black holes}
\label{sec:3}

\subsection{Preliminaries}
\label{sec:3.1}

According to
the discussion of the previous section, our method to find BH solutions
will consist in  embedding the
khronon field in an external BH metric. The latter satisfies 
the gravitational equations of motion in the vacuum. In the spherically
symmetric case we will start with the standard 
Schwarzschild metric in the Schwarzschild coordinates,
\be
\label{Schwarz}
\di s^2=\left(1-\frac{r_s}{r}\right)\di t^2-\frac{\di r^2}{1-\frac{r_s}{r}}
-r^2\di \Omega^2\;,
\ee
where
\be
\label{dO}
\di\Omega^2=\di\theta^2+\sin^2\theta \di\phi^2\;,
\ee
is the area element of the 2-sphere and $r_s$ is the Schwarzschild
radius. Using the Finkelstein coordinate 
\be
\label{vt}
v=t+r+r_s\ln{\bigg(\frac{r}{r_s}-1\bigg)}\;,
\ee
the previous metric can be cast into the form regular at 
the
Schwarzschild horizon,
\be
\label{Finkelstein}
\di s^2=\left(1-\frac{r_s}{r}\right)\di v^2-2\di v\di r
-r^2\di\Omega^2\;.
\ee

To simplify the khronon action, one observes that the curl of a hypersurface-orthogonal vector
(\ref{uphi}) identically vanishes,
\be
\omega_\m\equiv\epsilon_{\m\n\l\r}u^\n\nabla^\l u^\r=0\;.
\ee 
This implies 
\be
\label{oo}
0=\omega_\m\omega^\m=-\frac{1}{2}u_{\m\n}u^{\m\n}
+(u^\n\nabla_\n u_\m)^2\;,
\ee
where
\be
\label{Umn}
u_{\m\n}\equiv\d_\m u_\n-\d_\n u_\m\;.
\ee
Furthermore, 
\be
\label{bb}
\beta\int\di^4x \sqrt{-g} \nabla_\m u_\n\nabla^\n u^\m=\beta\int\di^4x \sqrt{-g} \big[(\nabla_\m u^\m)^2-R_{\m\n}u^\m u^\n\big]
\ee
where $R_{\m\n}$ is the background Ricci tensor. Combining (\ref{oo})
and (\ref{bb}) we 
obtain that
in the Schwarzschild 
geometry the khronon action can be written in the equivalent form
\be
\label{newAct}
S_\varphi=-M^2\int \di^4x\sqrt{-g}\bigg(\frac{\a}{4}u_{\m\n}u^{\m\n}
+\frac{\b+\l}{2}(\nabla_\m u^\m)^2\bigg)\;,
\ee
where we have used that the Ricci tensor for the
Schwarzschild metric vanishes.

The khronon equation of motion, following from the variation of
(\ref{newAct}) 
with respect to $\varphi$, has the form of a current
conservation,
\be
\label{khreq}
\nabla_\m J^\m=0\;,
\ee
with 
\be
\label{curr}
J^\m=\frac{P^\m_\n}{\sqrt X}\,\frac{\delta S_\varphi}{\delta u_\n}\;.
\ee
Here $X=\d_\m\varphi\nabla^\m\varphi$ and
\be
\label{P}
P^\m_\n=\delta^\m_\n-u^\m u_\n\;
\ee
is the projector on the hypersurface orthogonal to $u^\mu$; this
implies 
\be
\label{perpend}
u_\m J^\m=0\;
\ee
We are interested in static, spherically symmetric, 
asymptotically flat 
solutions of (\ref{khreq}). In the present context these notions
require some clarification.
The condition of staticity does not mean that the khronon field is
independent of time (indeed, we saw that in the case of the Minkowski
background $\varphi$ grows linearly with time). Rather it implies
that
all quantities invariant under the symmetry (\ref{reparam}) must be
constant in time. In particular, 
this applies to the components of
the vector $u_\m$. Combined with the requirement of spherical
symmetry, this implies that the only non-vanishing  
 components of the vector are $u_t$,
$u^r$ and that they depend only on the radial coordinate $r$.  
Note that we have chosen the position of the index '$t$' -- down and
'$r$' -- up so that the corresponding components of the vector coincide
in the Schwarzschild and Finkelstein frames:
\[
u_v=u_t~,~~~u^r\big|_{Finkelstein}=u^r\big|_{Schwarzschild}\;.
\]
For the
khronon field itself, staticity and spherical symmetry imply that, 
up to a reparameterization of the form
(\ref{reparam}), it can be cast into the form 
\be
\label{bhkh}
\varphi=t+f(r)\;,
\ee 
where $f(r)$ is a function of the radius.
Concerning the condition of the 
asymptotic
flatness, we will impose, in addition to the flatness of the metric at
$r\to\infty$, the requirement that the khronon field tends to the same
form as in Minkowski
space-time, $\varphi= t$. This implies the boundary conditions
at infinity:
\be
\label{boundc}
u_t\to 1~,~~~ u^r\to 0~,~~~\sqrt X\to 1
~~~~~~~\text{at} ~~r\to\infty\;.
\ee 

We now prove that asymptotically flat static spherically symmetric 
solutions of (\ref{khreq})
satisfy the stronger equation
\be
\label{khreqq}
J^\m=0\;.
\ee 
Taking the first integral of Eq.~(\ref{khreq}) we obtain
\be
\label{JrA}
J^r=\frac{C_1}{r^2}\;,
\ee
where $C_1$ is an integration constant. On the other hand, the
explicit expression for $J^r$ in terms of $u_\m$ reads,
\be
\label{Jr}
J^r=\frac{1}{\sqrt X}\bigg[\a\, u^r u_t\bigg(\partial_r^2u_t+\frac{2}{r}\partial_r u_t\bigg)
-(\b+\l)u_t^2\partial_r\bigg(\partial_r{u^r}+\frac{2}{r}u^r\bigg)\bigg]\;,
\ee
where the components of the vector satisfy the unit norm constraint
\be
\label{constr}
u_t^2-(u^r)^2=1-\frac{r_s}{r}\;.
\ee
After substituting (\ref{JrA})  into
(\ref{Jr}) and expanding the r.h.s. of (\ref{Jr}) at large $r$, the solution of the
resulting linear differential
equation reads
\be
\label{asysol}
u^r=\frac{C_1}{2(\b+\l)}+\frac{C_2}{r^2}+C_3 r\;,
\ee
where $C_2$, $C_3$ are integration constants. We see that the only
solution compatible with the boundary conditions (\ref{boundc}) is
obtained for the choice $C_1=C_3=0$. This implies 
$J^r=0$. Finally, from (\ref{perpend}) one finds that $J^t$ is also
zero\footnote{Another derivation of the equivalence between
(\ref{khreq}) and (\ref{khreqq}) in the spherically symmetric case,
which is also applicable for non-static configuration, is
presented in \cite{Blas:2010hb}. However, it is based on
the assumption that all leaves of the constant khronon field that
foliate the space-time are regular and simply connected. 
This a priori assumption appears too restrictive in the case of BHs.}.

We point out that Eq.~(\ref{khreqq}) is the same as the equation of motion for the
Einstein-aether model. Indeed, the latter is obtained by the variation
of the action
(\ref{newAct}) with respect to the vector $u_\m$ imposing that the
variations of the vector components preserve
the constraint $u_\m u^\m=1$.  This leads to the appearance of the
projector $P^\m_\n$ in the expression (\ref{curr}) for $J^\m$. Together with the
fact that spherically symmetric solutions are always
hypersurface-orthogonal \cite{Jacobson:2010mx} this proves 
 that static spherically symmetric asymptotically flat solutions in
 the khronometric and 
Einstein-aether theories are equivalent.

\subsection{Solutions}

Our strategy to find the khronon configuration corresponding to a
BH will be as follows. We solve the equation $J^r=0$
for the vector $u_\m$, and afterwards reconstruct the corresponding khronon
field. Introducing the notations 
\be
\label{notations}
U\equiv u_t~,~~~ V\equiv u^r~,~~~ \xi\equiv\frac{r_s}{r},
\ee
equations (\ref{khreqq}) and (\ref{constr}) take the form,
\bseq
\label{eqsx}
\begin{align}
\label{eq1x}
&\frac{U''}{U}-c_\chi^2\frac{V''}{V}+\frac{2c^2_\chi}{\xi^2}=0\;,\\
\label{eq2x}
&U^2-V^2=1-\xi\;,
\end{align}
\eseq
where prime denotes differentiation with respect to $\xi$.  Note that in the new coordinate the position
of the Schwarzschild horizon is $\xi=1$, the region $\xi<1$ lies
outside this horizon, while $\xi>1$ corresponds to the black hole
interior. 
Expressing $V$ from\footnote{The choice of the minus sign in
  front of the square root corresponds to the configuration where the foliation is Ê`infalling', which is
what one expects for the
   black hole
  configuration. The plus sign would be relevant for the white hole case.}  (\ref{eq2x})
\be
\label{Ur}
V=-\sqrt{U^2-1+\xi}
\ee 
we obtain a single equation for the $U$ component:
\be
\label{eqsolv1}
U''+\frac{c^2_\chi U}{U^2(1-c^2_\chi)-1+\xi}
\bigg[-(U')^2+\frac{(UU'+1/2)^2}{U^2-1+\xi}+\frac{2(U^2-1+\xi)}{\xi^2}\bigg]=0\;.
\ee

One observes that the denominator in the second term vanishes at the
point $\xi_c$ where
\be
\label{chor}
(1-c_\chi^2)U^2(\xi_c)=1-\xi_c\;.
\ee
This is nothing but the equation determining the position of the sound
horizon for the propagating khronon mode. Indeed, we saw in
Sec.~\ref{sec:2} that in Minkowski space-time with the vector $u_\m$
aligned along the time direction the wave component of the khronon
propagates along the
rays $|{\bf x}|=c_\chi t$. Denoting the tangent vectors to these rays
by $n_\mu$ we find that they satisfy
\be
\label{cone}
n_\m n^\m=(1-c^2_\chi)(u_\m n^\m)^2\;.
\ee 
This equation has covariant form and straightforwardly generalizes to curved backgrounds with generic vector $u_\m$
 where it describes the sound cone for the khronon waves. The defining
property of the sound horizon is that inside it all future-directed
khronon rays point towards the center, $n^r<0$ for any $n^\m$
satisfying (\ref{cone}) and $n^v>0$, while outside of it the sign of $n^r$
can be positive. This implies that {\em at} the sound horizon the sound
cone contains the vector  
$n^\m=(1,0,0,0)$. Substituting this into (\ref{cone}) and using the
metric (\ref{Finkelstein}) we obtain (\ref{chor}). 
Clearly, for the subluminal khronon,
$c_\chi<1$, the sound horizon lies outside the Schwarzschild horizon, 
$\xi_c<1$, while for the superluminal
case
$c_\chi>1$, the situation is opposite, $\xi_c>1$; 
for $c_\chi=1 $ both horizons
coincide. 

We are interested in khronon embeddings with a regular future sound
horizon\footnote{Let us mention that in the superluminal case
  $c_\chi>1$ one can also find solutions that are regular everywhere
  except at $r=0$ ($\xi=\infty$) and do not possess the sound horizon at
  all. However, the numerical study of spherically symmetric collapse
  in the Einstein-aether model 
 indicates that it always leads to the formation
 of a sound horizon  \cite{Garfinkle:2007bk}. Given the equivalence of the two theories
in the spherically symmetric setting, one expects this to hold in the
khronometric model as well.} that
are expected to appear as the result of a gravitational 
collapse. 
The regularity of the solution at the sound horizon implies that the
term within the square brackets in (\ref{eqsolv1}) must vanish at
$\xi_c$. This imposes a relation between the first derivative $U'$ and
the function $U$ itself at this point, providing an additional
boundary condition. Using (\ref{chor}) to simplify the formulas we obtain,
\be
\label{Uprimec}
U'(\xi_c)=\frac{1}{2(1-c_\chi^2)U(\xi_c)}\bigg[
-1+c_\chi\sqrt{1-\frac{8c_\chi^2(1-c_\chi^2)U^4(\xi_c)}{\xi_c^2}}\bigg]\;,
\ee 
where we have chosen the branch of the square root that gives a regular
result at $c_\chi=1$. 

We solved Eq.~(\ref{eqsolv1}) numerically
imposing the boundary
conditions (\ref{chor}), (\ref{Uprimec}) and
\be
\label{atzero}
U(\xi=0)=1.
\ee
This was done using a shooting procedure where the shooting parameter is
taken to be the value of the field at the sound horizon,
$U(\xi_c)$. This determines the positions $\xi_c$ of the
sound horizon and the first derivative of the field there through
(\ref{chor}), (\ref{Uprimec}). Given $U(\xi_c)$ and $U'(\xi_c)$ we
integrate Eq.~(\ref{eqsolv1}) from\footnote{In the numerical
  implementation of the algorithm one has to set the initial data
  slightly away from $\xi_c$ to avoid computational instabilities.} 
$\xi_c$ towards $\xi=0$ to find $U(0)$. Iterating this procedure we find the value $U(\xi_c)$
that allows to satisfy (\ref{atzero}). Finally, we extend the solution
inside the sound horizon by solving Eq.~(\ref{eqsolv1}) from $\xi_c$
to $\xi\to\infty$. In this way we obtain a unique solution for every
value of $c_\chi$.

\begin{figure}[!htb]
\begin{center}
\includegraphics[width=0.64\textwidth]{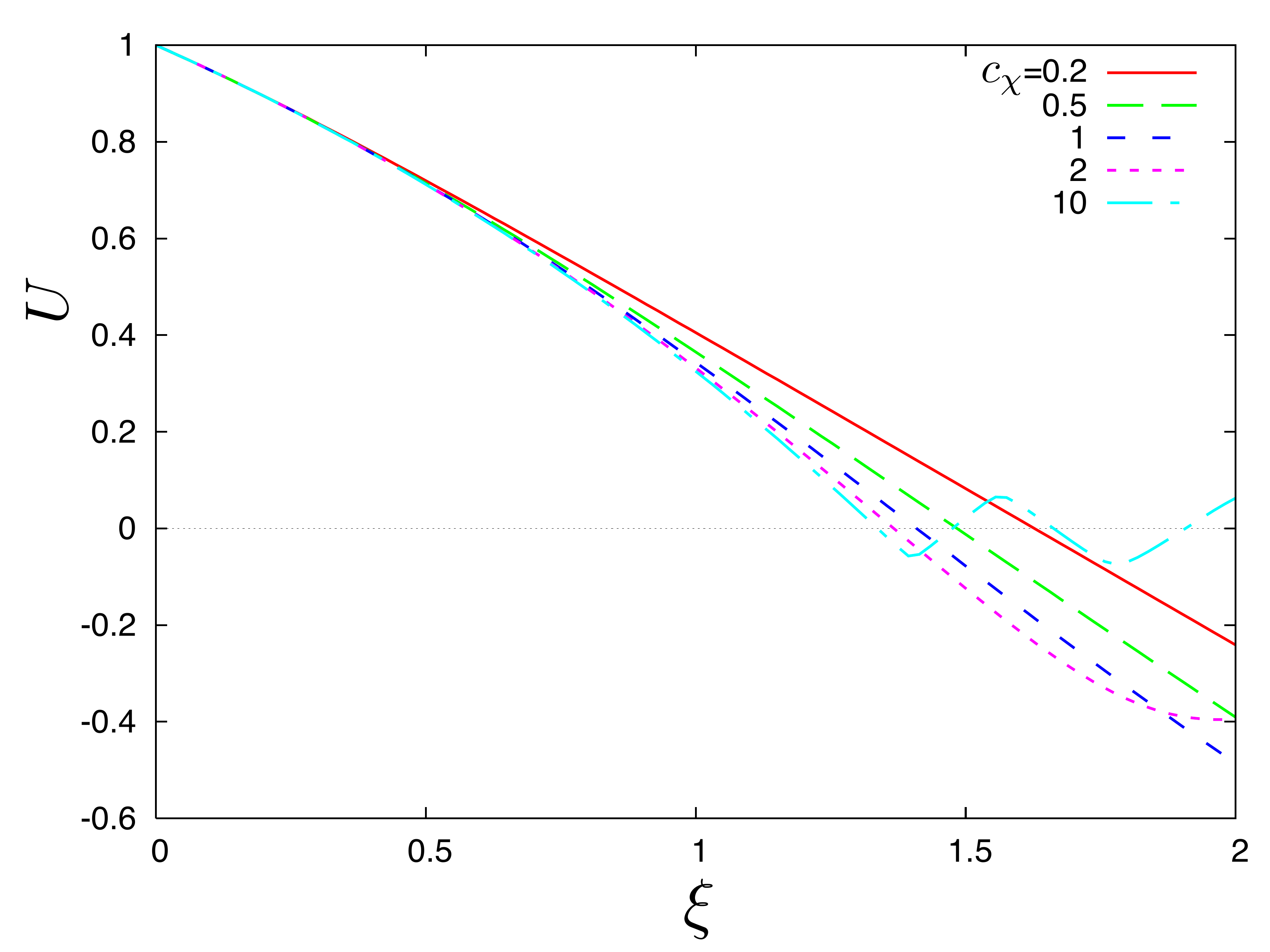}
\caption{The time component $U\equiv u_t$ of the vector $u_\m$ on the
  BH solution for several values of the khronon sound speed; the
  coordinate $\xi$ is related to the radius as $\xi\equiv r_s/r$.
\label{Fig:1}
}
\end{center}
\end{figure}
     
The resulting profiles $U(\xi)$ are plotted in Fig.~\ref{Fig:1} for
several values of the khronon sound speed $c_\chi$.
A few characteristics of the solutions are listed in
Table~\ref{tab:1}.
An important property, that is immediately clear from
Fig.~\ref{Fig:1}, is the existence of roots of the 
function $U(\xi)$ (i.e. points where $U(\xi)$ crosses zero). In  
Appendix~\ref{App:A} the existence of at least one root is proven
analytically for any value of  
$c_\chi$. 
Moreover, the numerical analysis indicates that the number of roots is
actually infinite with the function $U(\xi)$ exhibiting an oscillatory
behavior around the zero axis.
(The beginning of the oscillations is clearly visible in
Fig.~\ref{Fig:1} for the curve corresponding to $c_\chi=10$; for
smaller values of $c_\chi$ the oscillations lie outside the range shown
in the figure.) This is in agreement with the findings of
\cite{Ted}. 

Let us denote the smallest root of $U(\xi)$ by $\xi_\star$. Its values for
the numerical solutions at different $c_\chi$ are listed in
Table.~\ref{tab:1}. One observes that it mildly depends on the sound
speed and satisfies $\xi_\star>1$, $\xi_\star>\xi_c$. One can show that
$\xi_\star$ varies from $2$ for $c_\chi\to 0$ to  $4/3$ when
$c_\chi\to\infty$, see Appendix~\ref{App:A}. In the three-dimensional
space the point $\xi_\star$ corresponds to  a two-sphere lying inside 
the Schwarzschild and sound horizons. We are going to see that this
sphere plays a special role in the causal structure of the
khronometric BHs. 

The last quantity presented in Table~\ref{tab:1} is the
derivative of the function $U$ at $\xi_\star$,
$
U'_\star\equiv U'(\xi_\star).
$
These numerical data will be used below.
\begin{table}
\begin{center}
\begin{tabular}{|c|c|c|c|}
\hline
$c_\chi$ & $\xi_c$ & $\xi_\star$ & $U'_\star$\\\hline
0.2 & 0.52 & 1.63 & -0.647 \\\hline
0.5 & 0.81 & 1.48 & -0.764 \\\hline
0.75 & 0.93 & 1.44 & -0.819 \\\hline
1 & 1 & 1.41 & -0.858\\\hline
1.5 & 1.09 & 1.38 & -0.906 \\\hline
2 & 1.14 & 1.37 & -0.934 \\\hline
10 & 1.28 & 1.337  & -1.032 \\\hline
100 & 1.328 & 1.333 & -1.058 \\\hline
\end{tabular}
\end{center}
\caption{Several characteristics of the black hole
  solutions for different values of $c_\chi$. See the text for
  definitions.}
\label{tab:1}
\end{table}

\subsection{The universal horizon}

\begin{figure}[!htb]
\begin{center}
\includegraphics[width=0.6\textwidth]{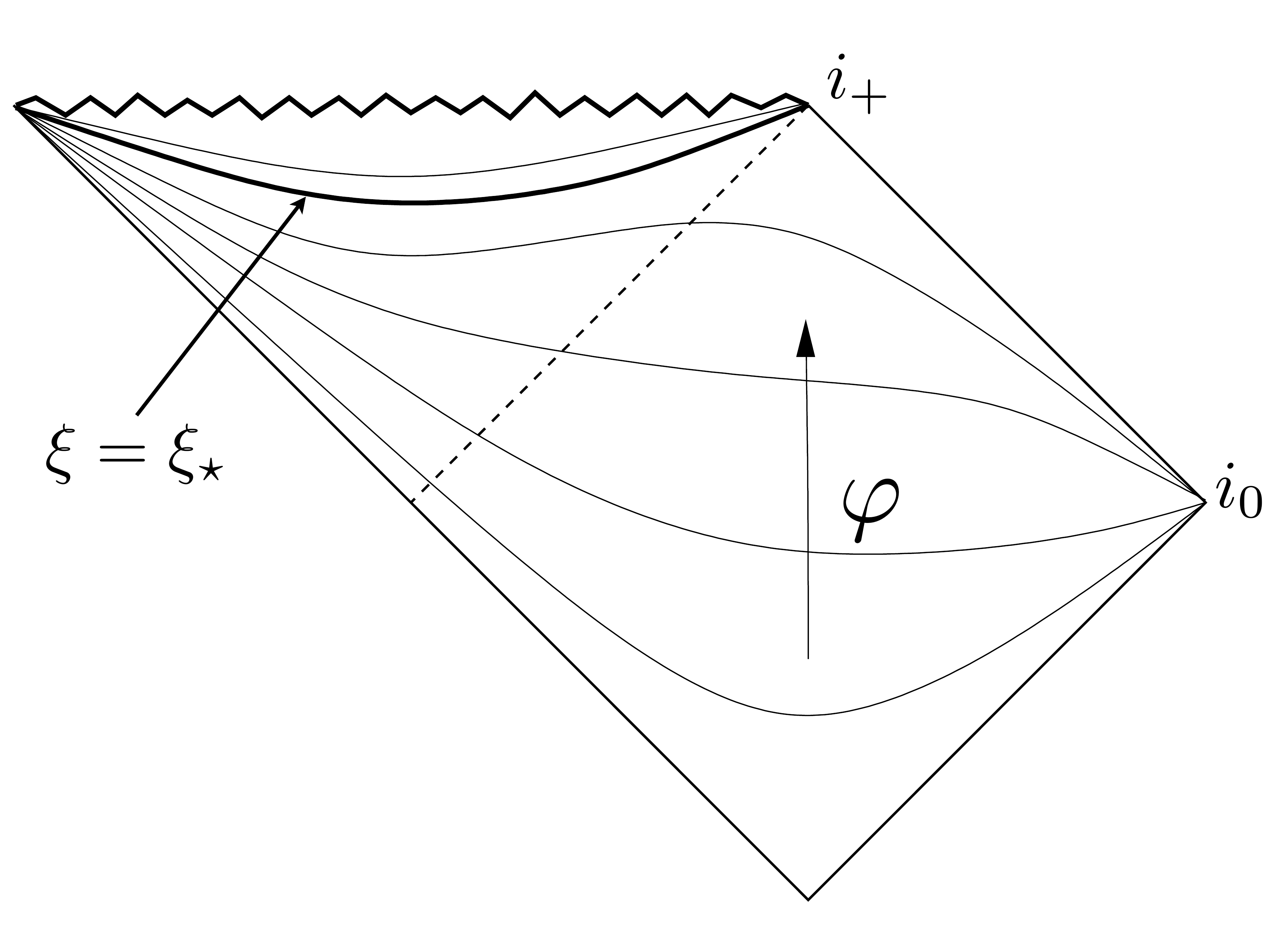}
\caption{The leaves of constant khronon field (thin solid lines)
  superimposed on the 
  upper half of the 
  Penrose diagram of the Schwarzschild black hole. The thick solid
  line shows the universal horizon.
\label{Fig:2}
}
\end{center}
\end{figure}

Let us now study the khronon configuration
corresponding to the BH solution. Taking the Ansatz
\be
\label{khsol}
\varphi=v+\varphi_1(\xi)\;
\ee
and recalling the definitions (\ref{uphi}), (\ref{notations}) we 
obtain the equation 
\be
\label{phi1eq}
\varphi'_1=\frac{u_\xi}{u_v}=\frac{-V(\xi)-U(\xi)}{\xi^2(\xi-1)U(\xi)}\;,
\ee
where, 
to avoid proliferation of multiplicative factors $r_s$, we have set $r_s=1$; this convention will be adopted from
now on.
It is straightforward to obtain the solution in the vicinity of $\xi=\xi_\star$.
Expanding the r.h.s. of (\ref{phi1eq})  we obtain ,
\be
\label{varphistar}
\varphi_1(\xi)\approx
\frac{1}{\xi_\star^2U'_\star\sqrt{\xi_\star-1}}\log{(\xi_\star-\xi)}~~~~~
\mathrm{at}~\xi\approx \xi_\star\;.
\ee
We observe that the khronon field logarithmically diverges at
$\xi_\star$. This implies that the leaves of constant khronon
coming from the spatial infinity accumulate at $\xi_\star$. None of them
penetrate inside the region $\xi>\xi_\star$. In other words, the leaves
foliating the interior of the sphere $\xi=\xi_\star$ are disconnected from
the spatial infinity. This fact is illustrated in  
Fig.~\ref{Fig:2}
where the khronon foliation is superimposed on the part of the BH 
Penrose diagram covered by the Finkelstein coordinates. 
The situation is similar to 
what happens in  GR for solutions containing Cauchy horizons. In both cases, the maximally extended
solution contains more than one connected (asymptotic) region where boundary conditions should be specified.
In the khronometric case, to know the solution at $\xi\geq \xi_\star$, one needs to specify the boundary conditions for the instantaneous mode
not only at $i_0$, but also at $i_+$ (see Fig.~\ref{Fig:2}). 

On the other hand, the sphere $\xi=\xi_\star$ simultaneously plays the
role of the universal causal horizon. Indeed,  
the khronon field sets the global time in the model at
hand. All signals, no matter how fast, can propagate only forward in
this global time. In this way the configuration of the khronon
determines the causal structure of space-time in Ho\v rava
gravity. From 
Fig.~\ref{Fig:2} it is clear that within this causal structure the
inner region $\xi>\xi_\star$ lies in the {\em future} with respect to the
outer part of the space-time. Thus no signal can escape from
inside the surface $\xi=\xi_\star$ to infinity (null asymptotic region between 
$i_+$ and  $i_0$) meaning that this surface is indeed a 
universal horizon, cf.~\cite{Ted}.

It should be pointed out that within the spherically symmetric
approximation that we have adopted so far the universal horizon is
regular, despite the apparent singularity (\ref{varphistar}) of the
khronon. Indeed, we have seen above that the field $u_\m$, which is
the proper invariant observable of the theory, is smooth at
$\xi=\xi_\star$. This implies that the singularity (\ref{varphistar}) can
be removed by the symmetry transformation of the form
(\ref{reparam}). It is easy to see that the transformation
\[
\varphi\mapsto \tilde\varphi=\exp \big[(\xi_\star^2U'_\star\sqrt{\xi_\star-1})\;\varphi\big]\;
\]
does the job: the redefined khronon field is analytic at $\xi_\star$. 
However, 
in the next section we will argue that the universal horizon
exhibits non-linear instability against {\em aspherical} perturbations of
the khronon field, which turn it into a physical singularity.  

\section{Beyond spherical symmetry: perturbations}
\label{sec:hair}

\subsection{Generalities}

The existence of the universal horizon seems to exclude the option (i)
for the resolution of the thermodynamical paradoxes mentioned in the
Introduction. However, this conclusion is premature: one has to
analyze the stability of the causal structure depicted in Fig.~\ref{Fig:2}
before making a definite statement. To this end we now study linear
perturbations of the khronon on top of the BH  
solutions found in the last section. This will also allow us to explore
the option (ii), namely 
possible existence of hair. As the background solution is not known in
 analytic form, we will characterize it  by functions
$U(\xi)$, $V(\xi)$ as defined in (\ref{notations}) and 
obeying Eqs.~(\ref{eqsx}).

The analysis of the khronon
perturbations must be performed in the preferred frame related to the
background khronon foliation. In this frame the equations for the
perturbations are second order in time, despite the presence of higher
spatial derivatives \cite{Blas:2009yd,Blas:2010hb,Jacobson:2010mx}.
Thus we introduce a new time coordinate $\tau$ that coincides
with the background khronon,
\be
\label{tauv}
\tau=v+\varphi_1(\xi)\;,
\ee
where $\varphi_1$ obeys Eq.~(\ref{phi1eq}). 
 In the new coordinates the
metric takes the form,\footnote{Recall that we work in the units with
  $r_s=1$.}  
\be
\label{taumetr}
\di s^2=(1-\xi)\di\tau^2 -\frac{2V}{\xi^2U}\di\tau
\di\xi-\frac{\di\xi^2}{\xi^4U^2}-\frac{\di\Omega^2}{\xi^2}\;.
\ee
Note that this metric is singular at $\xi=\xi_\star$ and thus only covers
the region outside the universal horizon. This is sufficient for our
purposes as we restrict to perturbations localized in this
region. 
The complete khronon field is
written as
\be
\label{taupert}
\varphi=\tau+\chi(\tau,\xi,\theta,\phi)\;,
\ee
where $\chi$ is a small perturbation. Due to the spherical symmetry
of the problem, different spherical harmonics of $\chi$ decouple from
each other at the linear level, which allows to 
consider the equations separately for
each multipole component $\chi_l$. The latter obeys the relation
\be
\label{lpole}
\Delta_{S^2}\chi_l=-L^2\chi_l\;,
\ee 
where $\Delta_{S^2}$ is the Laplacian on the unit 2-sphere and $L^2\equiv l(l+1)$.
To simplify notations we will omit 
the multipole label $l$ on $\chi$ in what follows.

Instead of directly linearizing the equations of motion, we find it
convenient to consider the quadratic action for the perturbations. 
Substituting (\ref{taumetr}), (\ref{taupert}) into 
(\ref{newAct}) after a straightforward (though somewhat lengthy)
calculation one obtains,
\be
\label{Squadr}
\begin{split}
S^{(2)}_\varphi=2\pi M^2\alpha\int \di\tau \di\xi\bigg[
U^2(\dot\chi')^2
&+\frac{L^2}{\xi^2}\dot\chi^2-2\xi^2U^3V\dot\chi'\chi''
-2L^2UV\dot\chi\chi'\\
&+A(\xi)(\chi'')^2+B_l(\xi)(\chi')^2+C_l(\xi)\chi^2\bigg]\;.
\end{split}
\ee
The coefficient functions $A$, $B_l$, $C_l$ are expressed in terms of
$U$, $V$ and their derivatives:
\bseq
\label{ABlCl}
\begin{align}
\label{A}
A(\xi)=&\xi^4U^4(V^2-c_\chi^2 U^2)\;,\\
\label{Bl}
B_l(\xi)=&-2\xi^4U^3V^2U''-\xi^4U^4VV''-2\xi^4U^2V^2{U'}^2
-2\xi^2U^4V^2-4\xi^4U^3U'VV'\\
&-4\xi^3U^4VV'-8\xi^3U^3U'V^2
+L^2\xi^2U^2V^2
-\xi^4U^5U''-2\xi^4U^3V^2U''\notag\\
&-c_\chi^2\big(-6\xi^4U^4{U'}^2-12\xi^3U^5U'
-3\xi^4U^5U''
+2 L^2\xi^2U^4+6\xi^2U^4V^2-3\xi^4U^4VV''\big)\;,\notag\\
\label{Cl}
C_l(\xi)=&-L^2\big[2\xi\, UU'V^2+\xi^2UV^2U''+2\xi^2UVU'V'+
\xi^2U^2{U'}^2+\xi^2U^3U''\\
&-c_\chi^2\big(-L^2U^2+\xi^2U^3U''+3\xi^2U^2{U'}^2
+10\xi U^3U'
+2U^4-2U^2V^2+\xi^2U^2VV''\big)\big]\notag\;.
\end{align}
\eseq
Note that the coefficient $A$ does
not depend on the multipole number $l$. In
deriving (\ref{Squadr}) and (\ref{ABlCl}) we made use of 
the background equations of
motion and integrated by parts to minimize the number of different
$\chi$-structures appearing in the action. Note that the action
(\ref{Squadr}) contains only two time-derivatives, as expected. 

Before analyzing the equation of motion following from (\ref{Squadr}) let us
discuss the boundary conditions that must be imposed on the
solutions. First, the field $\chi$ must vanish at the spatial
infinity, i.e. at $\xi\to 0$. Second, $\chi$ must be regular at the
sound horizon $\xi=\xi_c$. Note that the latter corresponds to a
zero of the coefficient $A(\xi)$. Determining the correct boundary
conditions at the universal horizon is slightly more complicated due
to the singularity of the metric (\ref{taumetr}) at $\xi_\star$. To bypass
this obstacle consider the linear
perturbations of the vector $u_\m$:  
\be
\label{Upert}
u^{(1)}_\tau=\xi^2U^2V\chi'~,~~~u^{(1)\,\xi}=-\xi^4 U^3\chi'~,~~~
u^{(1)}_\theta=U\d_\theta\chi~,~~~u^{(1)}_\phi=U\d_\phi\chi\;.
\ee
Here the positions of the upper and lower indices are chosen in
such a way that the presented components are invariant under the
coordinate change (\ref{tauv}). Thus they also define the vector
$u_\m$ in the Finkelstein frame. We will require that the
perturbations $u_\m^{(1)}$ in the latter frame are bounded at the
universal horizon. This is compatible with the assumption that the
black hole represents the end point of the gravitational collapse of a
smooth initial configuration. From (\ref{Upert}) we see that 
this requirement is equivalent to the condition that $\chi$
diverges at $\xi_\star$ not faster than $|\xi-\xi_\star|^{-1}$. We will see that
imposing this condition is enough for our purposes.

As a final remark we note that
the
expressions (\ref{ABlCl}) considerably simplify in the case
$c_\chi=1$, 
\bseq
\label{ABlCl1}
\begin{align}
\label{A1}
A=&\xi^4(\xi-1) U^4\;,\\
\label{Bl1}
B_l=&2\xi^4(1-\xi)(U^2+V^2){V'}^2-2\xi^4(1-\xi)VV'
+8\xi^3(1-\xi)U^2VV'\notag\\
&-\frac{\xi^4}{2}V^2+\xi^4U^2
-4\xi^3(1-\xi)U^2-2\xi^3U^4+L^2 \xi^2U^2V^2-2L^2\xi^2U^4\;,\\
\label{Cl1}
C_l=&-L^2\big[(2\xi V^2-10\xi U^2+\xi^2)UU'-2U^4+L^2U^2\big]\;,
\end{align}
\eseq
where we have  used  the background equations (\ref{eqsx}) to simplify the final expressions.

\subsection{Static perturbations: absence of hair}

Now we can prove that the khronometric BHs 
do not possess long range
hair\footnote{Of course, as in GR, BHs can have hair
  corresponding to  angular momentum and possible gauge charges. We
do not consider this standard hair, concentrating on those properties
of BHs that are peculiar to the khronometric model.}. At 
linear level such hair would manifest itself in the form of regular static
perturbations of the khronon field (cf.~\cite{Dubovsky:2007zi}). 
These obey the equation 
\be
\label{eq2}
(A(\xi)\chi'')''-(B_l(\xi)\chi')'+C_l(\xi)\chi=0\;.
\ee
Let us count the number of free parameters of the
general solution of Eq.~(\ref{eq2}) and compare
it with the number of boundary conditions. Consider the asymptotics of
the equation at spatial infinity, $\xi\to 0$. 
Keeping the leading
terms in the functions $A,B_l,C_l$, cf. (\ref{ABlCl}), 
we obtain,
\be
\label{eqsmallx}
-(\xi^4\chi'')''+2L^2(\xi^2\chi')'+L^2(2-L^2)\chi=0\;.
\ee
We will  look for solutions of the power-law form,
$\chi \propto \xi^\delta$. 
Substituting this Ansatz into (\ref{eqsmallx}) and solving the resulting
algebraic equation for $\delta$ one finds four roots:
\be
\label{deltas}
\delta_1=l-1~,~~~\delta_2=-l~,~~~\delta_3=l+1~,~~~\delta_4=-l-2\;.
\ee
Note that for $l=0$ we recover the asymptotics 
(\ref{asysol}). For $l\geq 1$ the solutions corresponding to 
$\delta_{2,4}$ grow at infinity and
must be rejected. Thus we are left with a two-parameter family of
decaying solutions\footnote{There is a subtlety in the case $l=1$. The
mode corresponding to $\delta_1$ is asymptotically constant and hence
is sensitive to non-linear corrections. In principle, these
corrections can make it diverge at $\xi\to 0$. Whether this happens or
not, is irrelevant for our argument.},
\be
\label{twop}
\chi\sim\chi_1\xi^{l-1}+\chi_3\xi^{l+1}\;.
\ee
These are further constrained
by the boundary conditions at the universal and sound
horizons. Expanding Eq.~(\ref{eq2}) at $\xi_\star$ we obtain
\be
\label{eqx*}
(\xi_\star^4(\xi_\star-1){U'_\star}^4(\xi-\xi_\star)^4\chi'')''
+\Big((\xi_\star-1)\xi_\star^2{U_\star'}^2(2\xi_\star^2{U_\star'}^2
-L^2)(\xi-\xi_\star)^2\chi' \Big)'=0\;.
\ee
Substituting the power-law Ansatz
\be
\chi\propto (\xi-\xi_\star)^\gamma
\ee
we obtain for the exponent
\be
\label{gammas}
\gamma_1=0~,~~~\gamma_2=-1~,~~~
\gamma_{\pm}=-\frac{1}{2}\pm
\sqrt{\frac{1}{4}+\frac{L^2}{\xi_\star^2 {U'_\star}^2}}\;.
\ee
Recalling that $\chi$ must grow not faster than $|\xi-\xi_\star|^{-1}$ we
obtain that the solution corresponding to $\gamma_-$ is excluded. This
gives one equation on the two parameters $\chi_1$, $\chi_3$. One more
equation follows from the requirement of regularity at the sound
horizon $\xi=\xi_c$. Thus in total we have two equations for two
parameters. Assuming that this system is not degenerate we conclude
that the unique solution is $\chi_1=\chi_3=0$ implying the absence of
hair. 

The fact that the equations following from the boundary conditions are
indeed non-degenerate can be explicitly verified in the case of high
multipoles, $L^2\gg 1$. For the sake of the argument,
we restrict 
to the case $c_\chi=1$, where the coefficients $A, B_l, C_l$
simplify, see Eqs.~(\ref{ABlCl1}).
Keeping only the leading
terms in $L^2$ in Eq.~(\ref{eq2}) we obtain,
\be
\label{eqlargeD}
(\xi^4U^4(\xi-1)\chi'')''+L^2\Big((-\xi^2U^2V^2+2\xi^2U^4)\chi'\Big)'-L^4U^2\chi=0\;.
\ee 
The form of this equation suggests to use the WKB method.
Thus we search for the solutions using the expansion,
\be
\label{WKB}
\chi(\xi)=\exp[L \,\s(\xi)+\ldots]\;.
\ee
Substituting this into (\ref{eqlargeD}) and restricting to the leading
order $O(L^4)$ we get,
\be
\xi^4U^2(\xi-1)(\s')^4+(-\xi^2V^2+2\xi^2U^2)(\s')^2-1=0\;.
\ee
This yields four solutions: 
\begin{align}
\label{case1}
&\chi_{1,\pm}\propto\exp\bigg[\pm L\int\frac{d\xi}{\xi U}\bigg]\;,\\
\label{case2}
&\chi_{2,\pm}\propto
\exp\bigg[\pm L\int\frac{d\xi}{\xi\sqrt{1-\xi}}\bigg]\;.
\end{align}
Note that $\chi_{1,\pm}$ are automatically regular
at the sound horizon $\xi_c=1$. We will see below that these solutions
describe the instantaneous khronon mode in the black hole background.
However, in the absence of sources 
they blow up either at spatial infinity or at the universal horizon 
(recall that $U'_\star <0$) in a way incompatible with the desired
boundary conditions presented above. Thus they do not
describe a valid static configuration.
  
The two remaining 
solutions are at first sight 
singular at the sound horizon $\xi=1$. However,  it is
straightforward to check that the 
adiabaticity condition
\be
\label{WKB_ap}
\frac{\s''}{L (\s')^2}\ll 1\;,
\ee  
is violated at $\xi\to 1$, which means that
 the WKB
approximation cannot be trusted in the vicinity of $\xi=1$.
To work out the constraints imposed by the regularity at the sound
horizon we have to solve Eq.~(\ref{eqlargeD}) explicitly at
$\xi\approx 1$ and then match the solution to the WKB form 
(\ref{case2}) in a region where both approximations hold. In the vicinity of the sound
horizon Eq.~(\ref{eqlargeD}) takes the form,
\be
\label{Eq_hori}
U_c^4(\xi-1)\chi''''+2U_c^4\chi'''+L^2U_c^4\chi''
+L^2 (2U_c^4-U_c^2+4U_c^3 U'_c)\chi'-L^4U_c^2\chi=0\;,
\ee
where we have expanded all the coefficients at $\xi=1$ and introduced
the notations, $U_c\equiv U(1)$, $U'_c\equiv U'(1)$.
After introducing the new variable 
\be
\zeta\equiv L^2 (\xi-1)\;
\ee
Eq.~(\ref{Eq_hori}) becomes to the leading order 
$O(L^{6})$,
\be
\zeta\frac{\di^4\chi}{\di\zeta^4}+2\frac{\di^3\chi}{\di\zeta^3}
+\frac{\di^2\chi}{\di\zeta^2}=0\;.
\ee 
The regular solution of this equation has the form
\be
\label{chi2x1}
\chi\propto \int^\zeta \di\zeta'\int^{\zeta'}\di\zeta'' 
~\frac{J_1(2\sqrt {\zeta''})}{\sqrt
  {\zeta''}}\;,
\ee
where $J_1$ is the Bessel function. The solutions 
(\ref{chi2x1}) and (\ref{case2}) must be matched at $L^2\ll|\zeta|\ll
1$. Using the asymptotics of $J_1$ one finds that at $\zeta<0$ ($\xi<1$), the solution (\ref{chi2x1})
contains both $\chi_{2,+}$ and $\chi_{2,-}$ components. The latter
diverges at spatial infinity implying that this solution also cannot
represent a static hair.

\subsection{Time-dependent perturbations: non-analyticity 
at the universal horizon}

We now return to the general case of time dependent
perturbations. We will concentrate on large multipoles where it is
possible to obtain approximate solutions using the WKB method. For
clarity we will also restrict to the case $c_\chi=1$; the reader can
easily get convinced that this restriction is not essential and that our
qualitative results hold for arbitrary $c_\chi$. Expanding the
equation following from (\ref{Squadr}) to leading order in $L$
and substituting $\chi$ in the form of a periodic function of time,  
\be
\chi\propto \e^{-i\omega \tau}\;,
\ee
we obtain,
\be
\label{eqlargeDomega}
\begin{split}
(\xi^4U^4(\xi-1)\chi'')''+i\omega(\xi^2U^3V\chi'')'
+i\omega(\xi^2U^3V\chi')''-i\omega L^2(UV\chi)'-i\omega L^2UV\chi'\\
+\big(\big[2L^2\xi^2U^4-L^2\xi^2U^2V^2-\omega^2U^2\big]\chi'\big)'
-\bigg(L^4U^2-\frac{\omega^2L^2}{\xi^2}\bigg)\chi=0\;.
\end{split}
\ee 
Upon substitution of the WKB Ansatz (\ref{WKB}) this takes the form,
\be
\label{sigmaOmega}
\begin{split}
\xi^4U^4(\xi-1)(\sigma')^4+2i\Omega \xi^2U^3V(\s')^3
&-2i\Omega UV\s'\\
-(\xi^2U^2V^2-2\xi^2U^4+\Omega^2U^2)(\s')^2
&-\bigg(U^2-\frac{\Omega^2}{\xi^2}\bigg)=0\;,
\end{split}
\ee
where we have introduced\footnote{This corresponds to focusing 
on frequencies of $O(L)$.}
\be
\label{Omega}
\Omega=\frac{\omega}{L}\;.
\ee
Remarkably,
for any value of $\Omega$ two of the solutions of
Eq.~(\ref{sigmaOmega}) are
\be
\label{sigmainst}
\s'_{1,\pm}=\pm\frac{1}{\xi U}\;.
\ee
They coincide with the expressions obtained in the static case in the same WKB limit,
cf.~(\ref{case1}). This implies that the radial dependence of these
modes is completely decoupled from their time-dependence, meaning that
the field changes simultaneously all over the space. Thus these modes
should be identified as mediating the instantaneous khronon
interaction. If there are no external sources these modes are put to zero by
the boundary conditions at spatial infinity and the universal
horizon. We will discuss what happens in the presence of sources in a
moment. 

Before that let us consider the two remaining roots of
Eq.~(\ref{sigmaOmega}), 
\be
\label{sigmapm}
\s'_{2,\pm}=\frac{-i\Omega}{\xi^2(\xi-1)}\bigg(\frac{V}{U}\pm
\sqrt{1+\frac{\xi^2(\xi-1)}{\Omega^2}}\bigg)\;.
\ee
Recalling that $V$ is negative and $V(1)=-U(1)$ we see that the branch
$\s_{2,+}$ is regular at the sound horizon $\xi=1$, while $\s_{2,-}$
has a logarithmic singularity. At infinity these modes correspond to
the infalling and outgoing waves respectively, and thus the previous singularity 
just expresses the standard result that no outgoing modes can escape
from the sound horizon. Both modes are regular
at the universal horizon and cross it in the inward direction. Indeed,
in the vicinity of $\xi_\star$ we write,
\be
\chi_{2,\pm}\propto\exp\big[-i\omega\tau+
L\,\s_{2,\pm}(\xi)\big]
\approx\exp\bigg[-i\omega\bigg(\tau-\frac{\log|\xi-\xi_\star|}{\xi_\star^2U_\star'
\sqrt{\xi_\star-1}}\bigg)\bigg]
=\e^{-i\omega v}\;,
\ee  
where in the last equality we have switched to the Finkelstein
coordinate $v$ using the relations (\ref{tauv}), 
(\ref{varphistar}). An implication of the above results is 
that the black hole is stable at the
linear level with respect to the high-multipole perturbations. Indeed,
an instability would manifest itself as a localized mode with positive
imaginary part of the frequency. However, it is straightforward to
check that for $\Im \omega>0$ the solutions $\chi_{2,\pm}$ blow up
either at $\xi=0$ or $\xi=1$, implying the absence of localized
modes. In this respect the situation is similar to the standard case
of the Schwarzschild black hole in GR \cite{Vishveshwara:1970cc,Price:1971fb,Wald:1979}.

One might try to 
extend the stability analysis to arbitrary multipoles including the
perturbations with $l\sim 1$. As we now discuss, this analysis 
appears unnecessary since there are
strong indications that the  
universal horizon is anyway destabilized at the
non-linear level due to the presence of the instantaneous interaction.
To show this, let us look closer at the structure of the
instantaneous mode. First, we notice that this mode can be separated
from the part of the signal propagating with finite velocity for
arbitrary multipoles provided that the frequency is large, $\omega\gg
1$. This is done by taking the limit $\omega\to \infty$ in the
equation following from (\ref{Squadr}) (alternatively, one can use
Eq.~(\ref{eqlargeDomega})) and extracting the leading
contribution. This gives the following equation for the radial
dependence of the instantaneous mode:
\be
\label{eqsimple}
(U^2\chi')'-\frac{L^2}{\xi^2}\chi=0\;.
\ee   
It is possible to show that this equation is equivalent to
\[
\D_{\perp}\chi=0\;,
\]
where 
\[
\D_\perp=\nabla_\m(g^{\m\n}-u^\m u^\n)\nabla_\n,
\]
is the spatial Laplacian along the surfaces of constant khronon. 
Importantly, we again observe that the time dependence of the
instantaneous signal completely factorizes from the spatial
dependence.  
The
general solution to Eq.~(\ref{eqsimple}) reproduces the asymptotics
corresponding to $\delta_2$, $\delta_3$ at $\xi\to 0$ and $\gamma_+$,
$\gamma_-$ at $\xi\to \xi_\star$, see Eqs.~(\ref{deltas}),
(\ref{gammas}). At high multipoles it reduces to the WKB solution
given by (\ref{sigmainst}). 

As already noted, without any sources the instantaneous mode vanishes
due to the boundary conditions. Let us study now what happens when there is a source for
the khronon perturbations outside the black hole. Physically, this can
be either an external field interacting with the khronon, or the
non-linear perturbations of the khronon  left after 
gravitational collapse. To be concrete, we consider the situation when
the source is located at a given distance from
the center.
Such a source produces instantaneous khronon perturbations
that fall off at infinity and at the universal horizon.
Due to the
factorization property
the perturbation in the
vicinity of $\xi_\star$ has the form,
\be
\label{nearhor}
\chi\propto |\xi-\xi_\star|^{\gamma_+} f(\tau)\;,
\ee
where $f(\tau)$ is the temporal profile of the source.
The key feature of this expression is that it is non-analytic at
$\xi=\xi_\star$. The   
singularity cannot be removed by a reparameterization of the khronon. 
The most straightforward way to see this is to 
consider the scalar invariants constructed from the vector $u_\m$.
Let us take as an example
\be
\label{square}
I\equiv u_{\m\n}u^{\m\n}\;,
\ee 
where $u_{\m\n}$ is defined in (\ref{Umn}). At large frequencies, but
still moderate multipoles, the leading linear contribution to this
object is 
\be
I^{(1)} \approx 4\xi^4UU'\dot{\chi}'\;.
\ee
Though $I^{(1)}$ itself is finite (actually, vanishing) at $\xi_\star$,
its derivatives of  high enough order diverge. Indeed, consider the
$l$th derivative of $I^{(1)}$ along the trajectory tangential to the
vector $u^\m$. Its leading coordinate dependence in the vicinity of the
universal horizon is 
\be
\label{td}
(u^\l\d_\l)^l I^{(1)} 
\propto |\xi-\xi_\star|^{\gamma_+-l} f^{(l)}(\tau)\;.
\ee
It is easy to check using Eq.~(\ref{gammas}) and the
data from the Table~\ref{tab:1} that $\gamma_+(l)<l$. In other words,
we obtain that for a given multipole $l$ the $l$th and higher derivatives of 
the quantity $I^{(1)}$
diverge. In particular, for the dipole $l=1$ the divergence occurs
already in the first derivative.
It is natural to expect that these divergences will show up at
non-linear orders of the perturbative expansion around the black hole
background turning the universal horizon into a singular surface.

\begin{figure}[!htb]
\begin{center}
\includegraphics[width=0.6\textwidth]{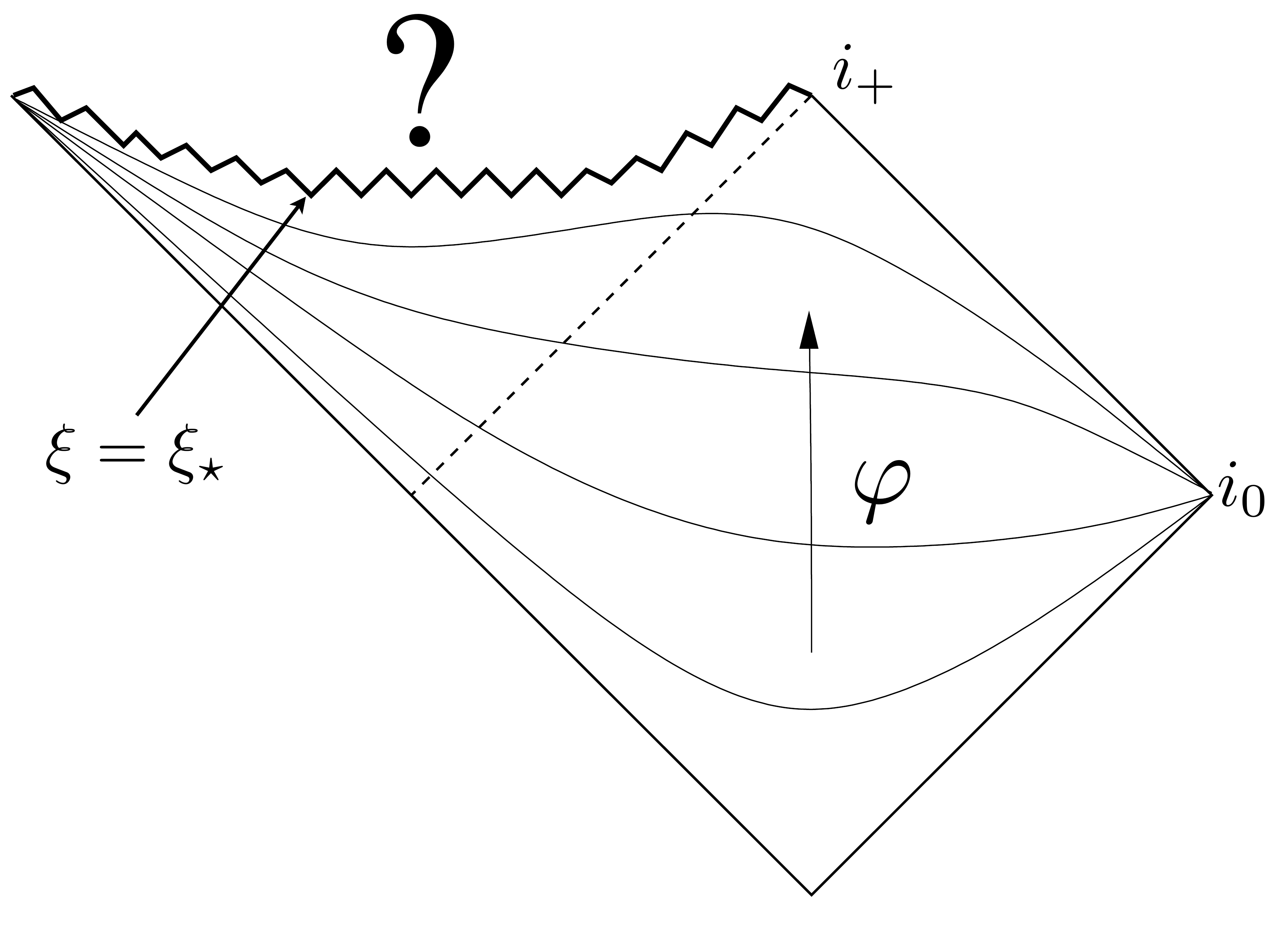}
\caption{The Penrose diagram of the terminal configuration of a
  realistic collapse in the khronometric theory.
The leaves of constant khronon field are shown by thin solid lines.
 Non-linear
  instabilities turn the universal horizon into a singular
  surface. The structure of the space-time inside this surface can be
  determined only within the full Ho\v rava gravity. 
\label{Fig:3}
}
\end{center}
\end{figure}

Consider now a realistic dynamical collapse producing a BH. This system will never be 
perfectly spherical and will leave behind perturbations of the metric and the khronon
present  in the region outside
the horizon and falling
down as a certain power of time \cite{Price:1971fb}. These perturbations
will source the instantaneous mode which in turn will produce the
divergences at the universal horizon. Thus we conclude that a realistic
collapse, instead of producing a regular universal horizon, results in
the formation of a finite area singularity at $\xi=\xi_\star$. The
Penrose diagram of the resulting configuration is depicted in 
Fig.~\ref{Fig:3}. 

The physical reason behind the instability of the universal horizon
can be understood as follows. 
As already stated, for the instantaneous
signal the universal horizon plays the role of a
Cauchy horizon. 
For an observer falling down the BH and crossing the universal horizon
the whole history of the universe outside the BH shrinks into a finite
time interval. This leads to an infinite blue-shift of signals sent
from the outside, these pile up at the universal horizon and turn it
into the singularity.
The situation is similar to the
instability of the inner Cauchy horizon in the case of
Reissner--Nordstr\" om and Kerr black holes in GR 
\cite{Matzner:1979zz,Poisson:1990eh}. However,
while in GR the signals destroying the Cauchy horizons are the standard
waves propagating along light cones, in the khronometric case they
must be instantaneous. 
Thus the presence of instantaneous interactions
plays the crucial role in the destabilization of the universal
horizon. Consequently, we expect this type of instability to be absent
in the case of the Einstein-aether theory where all modes propagate
with finite velocities.

The singularity that replaces the universal horizon is naked in the
sense that the complete determination of the khronon field requires
imposing some boundary conditions on it. However, for those observers
outside the black hole that are able to communicate only using
finite-velocity signals (like those interacting only with the ordinary
matter) the singularity is hidden by the corresponding sound
horizons. Thus, despite the presence of the singularity, the evolution of
the ordinary fields can still be unambiguously determined to the
extent that their interaction with the khronon can be neglected.

Finally, we do not know if the singularity discussed above is
associated  with a region of high space-time curvature, or if it is
just an irregularity of the khronon foliation. One can conjecture that in
the full Ho\v rava gravity the singularity is resolved by the higher
order terms, allowing for the complete determination of the
dynamics. Admittedly, whether this happens or not 
is an open issue requiring further study.    

\section{Discussion}
\label{sec:discussion}

Let us summarize the picture that emerges form our analysis and
speculate on its possible implications. It appears that in the healthy
Ho\v rava gravity a realistic BH formed as a result of a
gravitational collapse has a central region -- core -- where the
preferred foliation and possibly the metric are highly curved. The size
of the core is proportional to the Schwarzschild radius with the
proportionality coefficient mildly depending on the model
parameters. The core appears singular from the point of view of the
low-energy khronometric model. Optimistically, one may conjecture that
the singularity is resolved once the full structure of the healthy
Ho\v rava gravity with the inclusion of higher-derivative terms is
taken into account. 

The {\em surface} of the core can be probed from
outside by the high energy modes and by the low-energy instantaneous
interactions. In this sense the causal structure of the region outside
the core is trivial: the causal horizons appearing for certain modes in the low-energy
approximation are absent once the whole spectrum of the theory is considered. Without invoking the full-fledged Ho\v rava gravity it
is impossible to tell whether the triviality of the causal structure will
persist in the complete solution including the region {\em inside} the
core. We believe that this is plausible, because
 the existence of the BH core accessible from the
outside would eliminate the threat to BH thermodynamics raised
by the violation of Lorentz invariance. Indeed, when considering the
gedanken processes of \cite{Dubovsky:2006vk,Eling:2007qd} one
would also have to take into account the changes in the entropy of the
core: in the case of a trivial causal structure these changes are
detectable from outside. Though at the moment we cannot estimate the entropy of the core one can speculate that its increase
compensates for the decrease of the entropy in the part of the system
outside the BH and in this way the second law of
thermodynamics is saved.

The existence of the BH core can also have profound consequences on
the phenomenon of Hawking radiation. The (in)sensitivity of the latter
to trans-Planckian physics is a long-standing issue, see 
e.g. \cite{Corley:1996ar,Corley:1997pr,Unruh:2004zk,
Antonin:2011in,Barcelo:2008qe}. Under
broad conditions it 
has been shown that Hawking radiation is  determined by the low-energy
physics alone, 
in particular, by the properties of the sound horizons for low-frequency
modes
\cite{Corley:1996ar,Corley:1997pr,Unruh:2004zk,Antonin:2011in}. 
However,
a key assumption in these works is that the quantum fields are in the
vacuum with respect to the observers freely falling into the BH.
This assumption is likely to be
violated by the presence of the core just described: on the contrary,
it seems more natural to assume that the 
fields are in the vacuum with respect to the rest frame of the
core, which a priori is different from the free-falling one. It would
be very interesting to work out what impact this can have on the spectrum
of Hawking radiation. An extreme option would be that Hawking
radiation gets completely suppressed. Ho\v rava BHs would then
behave as stable objects: a kind of `dark stars'\footnote{This option
  does not exclude existence of a transient period just after the
  collapse when the BH would radiate nearly Hawking spectrum.}. 

The drastic modification of Hawking radiation is also suggested
by the following argument. 
Consider adding to the healthy Ho\v rava gravity a
field carrying a global charge.\footnote{We thank Sergei Dubovsky for
  suggesting to consider this setup.} 
Imagine that a  macroscopically large BH is formed and is endowed with large
global charge. It is well-known that if the BH evaporates in
the standard manner, the charge conservation will be grossly
violated. The immediate way to see this is to note that the standard
Hawking radiation is charge-symmetric and thus particles and
anti-particles are radiated in equal amounts leaving zero net charge
after the BH evaporation. Even allowing for modifications of
the radiation introducing a charge asymmetry will not save the day if
the radiation remains (approximately) thermal with the temperature of
the order of the standard Hawking temperature $T_H$. Indeed, the BH 
starts giving away the charge only when $T_H$ exceeds $m$, the
mass of the charged particles. Using the standard expression 
\[
T_H\sim M_P^2/M\;,
\]   
one estimates the BH mass at this moment
\[
M\sim M_P^2/m\;.
\]
Conservation of energy then implies that the charge that can be given
back by the BH is bounded above  
\cite{Dvali:2007hz,Dvali:2008tq} by
\[
Q\lesssim M_P^2/m^2\;.
\]
On the other hand, the initial charge of the BH can be as
large as 
\[
Q_0\sim M_0/m\;,
\]
where $M_0$ is the initial BH mass which, in its turn, can be
arbitrarily large. Note that one can further relax the assumption of
an approximate thermality of the radiation in the above argument. To show the
non-conservation of charge it is sufficient to assume that {\it (a)}
the BH evaporates and {\it (b)} that it cannot emit massive
particles during most of its evolution. 

The charge non-conservation would prevent Ho\v rava gravity from being
a weakly coupled model of quantum gravity. Indeed,
consider for concreteness the case when the charge is carried by a
scalar field with a global $U(1)$ symmetry. This symmetry is preserved at the
perturbative level. Thus any violation of the charge conservation can
stem only from non-perturbative gravitational effects. But in a
weakly coupled theory, these are expected to be exponentially
suppressed \cite{Kallosh:1995hi} meaning that the charge must be
conserved, at least with exponential accuracy. 

The only way to reconcile BH physics with the charge
conservation, and thus save Ho\v rava gravity as a candidate to a
consistent quantum theory, is to admit that either even
large BHs can emit massive charged particles, or that BHs 
do not evaporate at all. Clearly, both options would present
qualitative departures from the standard picture of Hawking
evaporation.

\paragraph*{Acknowledgments}

We are grateful to 
Eugeny Babichev, Sergei Dubovsky, Jaume Garriga, 
Dmitry Gorbunov, Ted Jacobson, Stefano Liberati, Shinji Mukohyama, 
Oriol Pujol\`as, Slava Rychkov, Thomas Sotiriou for illuminating
discussions. S.S. thanks the Theoretical Physics Group at IFAE,
Barcelona, and the Elementary Particle Theory Group at SISSA, Trieste,
for warm hospitality during various stages of this work. We
also thank the organizers and participants of the Peyresq 16 Meeting
for numerous stimulating conversations.  
This work was supported in part by the Swiss National Science Foundation, 
grant IZKOZ2\_138892/1 of the International Short Visits Program
(D.B.), 
the Grants of the President of Russian Federation
NS-5525.2010.2 and MK-3344.2011.2~(S.S.) and the RFBR grants
11-02-92108, 11-02-01528~(S.S.).

\appendix

\section{Absence of instantaneous propagation in Einstein-aether
  theory} 
\label{App:B}

We have seen in Sec.~\ref{sec:2} that the khronometric model exhibits
instantaneous interactions. From the mathematical perspective this is
due to the fact that the equation describing the khronon is of fourth order
because of two extra space-like derivatives. On the other hand,
the Einstein-aether theory is described by second-order hyperbolic
equations and one does not expect any instantaneous propagation in
this case. This seems puzzling as khronon can be viewed as just the
restriction of the aether to its longitudinal part. 
However, this is precisely the
restriction that leads to instantaneous signals: in this Appendix we
show that in the full aether theory the instantaneous piece is canceled by
the transverse modes. 

The aether Lagrangian has the form
\be
{\cal L}_{ae}=-\frac{M^2}{2}\Big[\a (u^\m\nabla_\m u_\n)^2 +
\a' \nabla_\m u_\n \nabla^\m u^\n
+\b \nabla_\m u^\n\nabla_\n u^\m + \l (\nabla_\m u^\m)^2\Big]\;.
\ee
This differs from the khronon Lagrangian (\ref{Act}) by the presence
of the term with the coefficient $\a'$. The vector $u_\m$ is subject
to the unit-norm constraint
\be
\label{unorm}
u_\m u^\m=1\;.
\ee
We are interested in the dynamics of perturbations in flat space-time
around the background aether configuration
\be
\bar u_0=1~,~~~~\bar u_i=0\;.
\ee
Due to the constraint (\ref{unorm}) only space-like components of the
aether are excited at the linear level. One separates the
perturbations into the longitudinal and transverse parts,
\be
u_i-\bar u_i=\d_i\chi+u^\perp_i~,~~~~\d_i u^\perp_i=0\;.
\ee
The longitudinal part is described by the quadratic khronon Lagrangian
(\ref{chiAct}), where one should make the substitution:
\bseq
\label{appsubst}
\begin{align}
\a&\mapsto \a+\a'\;,\\
\b+\l&\mapsto \b+\l+\a'\;.
\end{align}
\eseq
The Lagrangian for the transverse component is
\be
{\cal L}_{ae}^\perp=\frac{\a+\a'}{2}({\dot u}_i^\perp)^2
-\frac{\a'}{2}(\d_j u_i^\perp)^2\;.
\ee

We now introduce coupling of the aether field to the source
(\ref{Ssource}). The contribution of the longitudinal component to the
exchange amplitude is given by Eqs.~(\ref{ampl}), (\ref{Green}) (again
with the substitution (\ref{appsubst})), and thus contains an
instantaneous piece. However, a similar instantaneous contribution,
but with the opposite sign, is present in the amplitude describing
the exchange of the transverse mode:
\be
\label{Greenperp}
\begin{split}
G^{\perp\;ret}_{ij}(x)&=\int\frac{\di^4p}{(2\pi)^4}\;
\frac{\e^{-ipx}}{(\a+\a') p_0^2-\a'{\bf p^2}}\;
\bigg(\delta_{ij}-\frac{{\bf p}_i{\bf p_j}}{{\bf p}^2}\bigg)\\
&=-\frac{\theta(t)\delta(|{\bf x}|-c_\perp
  t)}{4\pi\sqrt{\a'(\a+\a')}|{\bf x}|}\,\bigg(\delta_{ij}-
\frac{{\bf x}_i{\bf x}_j}{|{\bf x}|^2}\bigg)
-\frac{\theta(t)\theta(|{\bf x}|-c_\perp t)}{4\pi(\a+\a')}\,
\left(\frac{3{\bf x}_i{\bf x}_j-\delta_{ij}|{\bf x}|^2}{|{\bf x}|^5}\right)t\;,
\end{split}
\ee
where 
$
c_\perp=\sqrt{\frac{\a'}{\a+\a'}}
$
is the velocity of the transverse mode. The last term in
(\ref{Greenperp}) cancels with that in (\ref{Green}) outside the wider sound cone, 
\[
|{\bf x}|=\max{(c_\perp,c_\chi)}\,t\;.
\]

\section{Spherical solutions: analytic results}
\label{App:A}

In this appendix we present a few analytic results about the solutions
of Eq.~(\ref{eqsolv1}) (or equivalently, the system
(\ref{eqsx})) with the boundary condition
(\ref{atzero}) at $\xi=0$. 

We begin by proving that if there is a point $\xi_c$
where (\ref{chor}) is satisfied, the solution will inevitably cross
zero. In other words, all solutions having a sound horizon also
possess a universal horizon. Let us assume the opposite. From
(\ref{Uprimec}) we observe that the derivative of $U(\xi)$ at the
sound horizon is negative,
\be
\label{negat}
U'(\xi_c)<0\;.
\ee
If $U(\xi)$ stays positive at $\xi>\xi_c$ at least one of the following
conditions must be satisfied:
\begin{itemize}
\item[(i)] there is a point $\xi_1>\xi_c$ where 
$U'$ first changes sign, i.e. $U'(\xi_1)=0$, $U''(\xi_1)>0$,
\item[(ii)] $U(\xi)$ tends to a non-negative constant with $U'(\xi)$ 
asymptotically approaching zero from below; this option implies that
$U''$ is positive at $\xi\to\infty$.
\end{itemize}
We proceed to show that none of these two option can be
realized. Consider the superluminal case $c_\chi\geq 1$. Then it is easy
to see that the
combination 
\[
F(\xi)\equiv U^2(\xi)(1-c^2_\chi)-1+\xi 
\]
is positive at $\xi_c<\xi<\xi_1$. Indeed, this combination vanishes at
the sound horizon, while its derivative
\[
F'=2(1-c^2_\chi)UU'+1
\]
is positive in the above interval. According to the previous
reasoning the factor in front of the square brackets in Eq.~(\ref{eqsolv1})
is positive at the point $\xi_1$, and it is straightforward to see that the quantity in the
square brackets is positive as well (recall that $U'(\xi_1)=0$). This
implies that $U''(\xi_1)$ is 
negative and we arrive at a contradiction. 

To exclude the option (ii) consider Eq.~(\ref{eqsolv1}) at large
$\xi$. For $U''$ to be positive, the combination in the square
brackets must be negative. This is possible only if
$
(U')^2> C^2/\xi\;,
$
where $C$ is some constant. This implies $U'< -C/\sqrt{\xi}$ and thus
$U$ cannot asymptote to a non-negative constant.

It remains to prove the statement in the subluminal case
$c_\chi<1$. The option (ii) is excluded by the same reasoning
as above because the combination $F(\xi)$ is positive at
$\xi\to\infty$. 
However, the option (i) is trickier as
it is not possible to argue that $F(\xi)$ is positive in the interval
$\xi_c<\xi<1$. Thus there may exist a turning point $\xi_1$ in this
interval. To rule out this possibility 
we
exploit the continuity of the solution in the parameter $c_\chi$. If
the universal horizon disappears for some values of $c_\chi$, there
must be a critical value $c_\chi^{(1)}$ such that $\xi_1$
exists for $c_\chi<c_\chi^{(1)}$ and does not exist
for $c_\chi>c_\chi^{(1)}$. For $c_\chi=c^{(1)}_\chi$ the derivative
$U'(\xi)$ touches zero at $\xi_1$,
$$
U'(\xi_1)=U''(\xi_1)=0\;.
$$
Then from Eq.~(\ref{eqsolv1}) we obtain $U(\xi_1)=0$. But this,
combined with the requirement $\xi_1\leq 1$, implies that the
$V$-component of the vector, Eq.~(\ref{Ur}), is
undefined in the vicinity of this point. We have again arrived at a
contradiction. This completes the proof.\\

The system (\ref{eqsx}) admits analytic solution in two limiting
cases: $c_\chi=0$ and $c_\chi\to\infty$. In the first case
Eq.~(\ref{eq1x}) degenerates into $U''=0$ implying that the solution
is a linear function,
\[
U=1-a\,\xi\;.
\]
The constant $a$ is fixed by considering the first approximation in
the small velocity $c_\chi\ll 1$ and imposing regularity at the sound
horizon. In this limit (\ref{Uprimec}) reduces to 
\[
2U'(\xi_c) U(\xi_c) =-1.
\]
Solving for $\xi_c$ from (\ref{chor}) in the same limit and substituting in the
previous equation one gets  $a=1/2$.
This yields the position of the universal horizon in this case,
\[
\xi_\star\to 2~~~~\mathrm{at}~~~~ c_\chi\to 0\;.
\]

On the opposite extreme $c_\chi\to\infty$ we can neglect the first
term in (\ref{eq1x}) and obtain the solution
\be
\label{solut}
V=b\,\xi^2~,~~~~
U^2=1-\xi+b^2\xi^4\;.
\ee
The solution exists at all $\xi$ only if $b\geq b_0$, where 
$b_0=\frac{\sqrt{27}}{16}$. In the limit $c_\chi\to\infty$ the condition
(\ref{chor}) for the position of the sound horizon becomes
\[
U(\xi_c)=0\;.
\]
For $b>b_0$ the functions $U(\xi)$ corresponding to 
(\ref{solut}) are strictly positive and thus the solutions 
do not possess a sound horizon. We believe that these solutions are
unphysical and cannot be formed in the gravitational collapse. The
physical solution that has a sound horizon corresponds
to\footnote{Strictly speaking, the limit of the physical solution at
  $c_\chi\to\infty$ is given by (\ref{solut}) only for
  $\xi\leq\xi_c$. At larger $\xi$ the numerical studies show that
  $\lim_{c_\chi\to\infty} U(\xi)\to 0$.} $b=b_0$. The
sound horizon in this case is located at $\xi_c=4/3$ and coincides
with the universal horizon.

\end{document}